# Imaging of high-Z material for nuclear contraband detection with a minimal prototype of a Muon Tomography station based on GEM detectors

Kondo Gnanvo[1*], Leonard V. Grasso III[1], Marcus Hohlmann[1], Judson B. Locke[1], Amilkar S. Quintero[1], Debasis Mitra[2]

[1] *Department of Physics and Space Sciences, Florida Institute of Technology, Melbourne, FL 32901, USA*
[2] *Department of Computer Sciences, Florida Institute of Technology, Melbourne, FL 32901, USA*

*Abstract* Muon Tomography based on the measurement of multiple scattering of atmospheric cosmic ray muons in matter is a promising technique for detecting heavily shielded high-Z radioactive materials (U, Pu) in cargo or vehicles. The technique uses the deflection of cosmic ray muons in matter to perform tomographic imaging of high-Z material inside a probed volume. A Muon Tomography Station (MTS) requires position-sensitive detectors with high spatial resolution for optimal tracking of incoming and outgoing cosmic ray muons. Micro Pattern Gaseous Detector (MPGD) technologies such as Gas Electron Multiplier (GEM) detectors are excellent candidates for this application. We have built and operated a minimal MTS prototype based on 30cm × 30cm GEM detectors for probing targets with various Z values inside the MTS volume. We report the first successful detection and imaging of medium-Z and high-Z targets of small volumes (~0.03 liters) using GEM-based Muon Tomography.

## 1. Introduction

Standard radiation detection techniques currently employed by portal monitors at international borders and ports are not very sensitive to radiation emanating from nuclear material if that material is well shielded. The idea of using cosmic ray muons for Muon Tomography (MT) based on the measurement of multiple scattering [1] of atmospheric cosmic ray muons as a promising technique to probe threat objects made of high-Z material, e.g. uranium or plutonium, and shielded nuclear material was originally proposed by a team at Los Alamos National Laboratory [2,3]. We propose the use of Gas Electron Multiplier (GEM) detectors [4] as the tracking devices for the Muon Tomography Station (MTS). GEM detectors are compact, have low mass, and can reach spatial resolutions down to about 50 μm. Results of Monte Carlo simulation studies on the performance expected from such a compact GEM-based MTS were reported previously [5]. In this paper we report the first successful reconstruction and imaging results using experimental data for targets with different Z values and shapes placed inside a minimal GEM-based MTS prototype.

## 2. Gas Electron Multiplier Detectors

### 2.1. Production and Assembly of GEM detectors

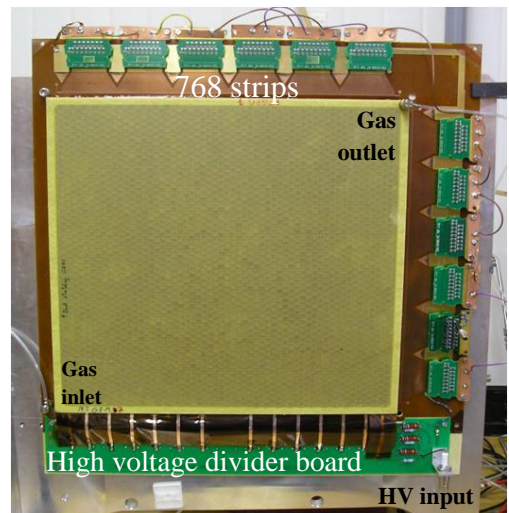

Figure 1: Triple-GEM Detector.

We built several 30 cm × 30 cm triple-GEM detectors for the first GEM-based MTS. Details of the GEM detector construction can be found in Ref. [6]. The design is based on the GEM detectors built by the TERA Foundation [7], which are in turn an upgraded version of the GEM detectors for the COMPASS experiment [8] at CERN. All detector components and the HV board were produced in the Electronics and PCB facilities at CERN. Systematic HV tests were performed before and during assembly to monitor the quality of the GEM foils. We assembled seven triple-GEM detectors and one double-GEM detector at CERN. Fig. 1 shows the picture of one of our triple-GEM detector on its stand for X-ray testing.

### 2.2. Commissioning

After the assembly, all seven triple-GEM detectors were tested under HV in 100% $CO_2$ to verify that there is no high leakage current from discharges caused by a short-circuit or by dust particles present in the chamber during the assembly process. $CO_2$ gas is used to avoid electron amplification in the chamber during this preliminary test. Six of the seven detectors passed the HV test and were consequently operated with an Ar/$CO_2$ 70:30 gas mixture on an X-ray test bench. At a total bias high voltage of 3.7 kV signal pulses became observable.

* Corresponding author: Tel.: +1 (321) 674-7339, Fax: +1 (321) 674-7482, e-mail: kgnanvo@fit.edu

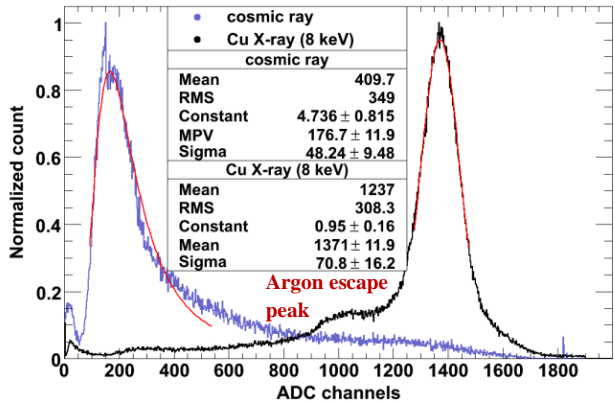

Figure 2: Pulse height spectra for a triple-GEM: Landau fit for cosmic ray muon spectrum and Gaussian fit for 8 keV Cu X-ray spectrum.

Fig.2 shows typical pulse height spectra obtained with one of the triple-GEM detectors exposed to a 8 keV Cu X-ray source. A typical Landau shape spectrum from cosmic ray data is also shown in Fig. 2 for data collected in 12 hours (~300,000 muons) with 1/6 of the total active detector area defined by 128-pins of the Panasonic connector used for the triple-GEM detector readout strips, ganged together to a single-channel amplifier. One of the triple-GEM detectors failed the HV test under $CO_2$ with a very high leakage current at a high voltage of 2 kV indicating a possible short circuit due to metallic contact. The problem is under investigation.

## 3. GEM-based Muon Tomography Station

### 3.1. MTS Geometry

We built a simple mechanical stand for the MTS that could accommodate several GEM detectors with 30 cm × 30 cm active area in each of the top and bottom tracking stations. For the first data run, a total of four triple-GEM detectors used in the MTS with two in the top station and two in the bottom station (see Fig. 3). The distance between the detectors was 9.45 cm as dictated by the size and routing of the readout cabling; the gap between the top and the bottom station was 10.1 cm. A thin press-board plate was inserted between the two tracking stations, at the center of the MTS (z=0) to support the targets. This setup, which we refer to as a "minimal" MTS, does not accommodate detectors on the side.

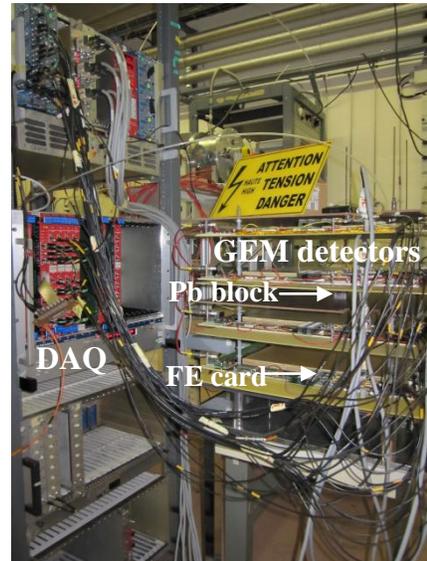

Figure 3: "Minimal" MTS with a Pb target and DAQ.

### 3.2. Front End Electronics and Trigger

For the readout of the x and y strips of the GEM detectors, we used eight Gassiplex front end (FE) cards [9]. Each FE card has 96 channels and uses 96-pin SAMTEC input connectors. The GEM detector readout board has 768 strips connected to six 128-pin Panasonic connectors in each x and y direction. With eight Gassiplex FE cards and four GEM detectors in the MTS, we could only instrument one Panasonic connector each in x and y per detector (2 × 128 strips readout), which means that we were able to read out a central area of 5 × 5 $cm^2$ for each of the GEM detectors. We produced an adapter PCB to interface the 96-pin SAMTEC connectors of the FE cards to the 128-pin Panasonic connectors on the GEM detectors. We needed to gang together pairs of strips for 64 of the 128 strips in order to match the 128 input strips on the detector to the 96 channels of the FE cards. For the external trigger, the coincidence signal of two 5 × 5 $cm^2$ plastic scintillators with PMT readout was used, with one scintillator placed just above the upper GEM of the top station and the other below the lower GEM of the bottom station. The two counters were carefully aligned with the active area of the GEM detectors in the MTS.

### 3.3. Data Acquisition System

The DAQ system, based on the CAST Micromegas detector DAQ [10], was composed of:

- a VME crate with a CAEN Controller card VME-MXI2 [11], four V550 CAEN Readout Analog Multiplexed Signal (CRAMS) modules [11] to digitize the analog signals from the FE cards, one V551 CAEN sequencer card [11] that receives the trigger signal, produces the control signals (Clock, Track/Hold, Clear) for the FE

- cards, receives a Data Ready signal from CRAMS, clears the CRAMS, resets the DAQ at the end of an event
- a NIM crate with a low voltage supply for the FE cards and a NIM-TTL-NIM level adapter to convert the trigger signal to a NIM signal for the DAQ and CAEN N470 [11] HV supply for the GEM detectors
- LabView software, upgraded from CAST DAQ software [10], which has an online component to control the VME crate, read and format the pedestal and event data, save the formatted data from into output data files and an offline component to decode the raw data, perform zero suppression, pedestal subtraction and display the data for a given run file. The offline software is also used to monitor the performance of the MTS. Fig. 4 shows a display of a raw cosmic muon event recorded by the MTS.

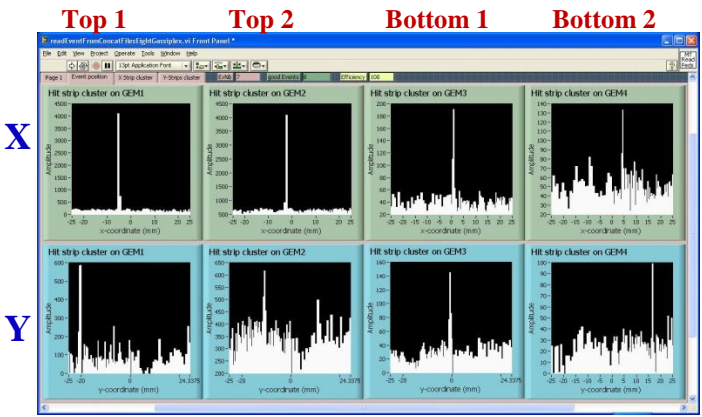

Figure 4: A cosmic muon recorded on both x (top) and y (bottom) strips with the 4 detectors displayed with the Labview DAQ software.

## 4. Detection and Imaging with a minimal GEM-MTS

### 4.1. Cosmic ray data runs

We performed a month-long cosmic ray muon run at CERN in April 2010 with the minimal MTS setup described above, i.e. for an MTS volume defined by a $5 \times 5$ cm$^2$ active area of the GEM detectors and a 10.1 cm gap between the top and bottom tracking station. We took data for four different scenarios at a trigger rate of about 1,000 events per day. This rate given by the small solid angle defined by the two $5 \times 5$ cm$^2$ trigger counters is consistent with what we expect from a Monte Carlo simulation. Cosmic data were taken for the following MTS scenarios:

- The first run was performed with an empty MTS volume for 2 days (~1,900 events) to evaluate the alignment of the four GEM detectors and the performance of the MTS.
- The second run was performed for 3 days (~3,000 events), with a $30 \times 30 \times 30$ mm$^3$ iron (Fe, Z=26) block on a target plate located at z = 0 in the MTS volume.
- The third run was performed for 3.5 days (~3,600 events) with a lead (Pb, Z=82) target block with dimensions $28 \times 20 \times 30$ mm$^3$ along the x, y, z axes of the station, respectively.
- Finally, a 5-day run (~5,000 events) with a 30 mm diameter Tantalum (Ta, Z=73) cylinder of 16 mm height.

### 4.2. Reconstruction of the various scenarios

We ran a point-of-closest-approach (POCA) algorithm [12] on the data to reconstruct the scenarios described in section 4.1. Fig. 5 displays the 3D reconstruction where each point represents the reconstructed "interaction point" of the muon deflected by the target. The shading (color) represents the magnitude of the deflection angle. Except for the empty scenario case, the high-angle POCA "points" are correctly reconstructed mainly at the target locations, especially for high-Z materials like Pb and Ta targets.

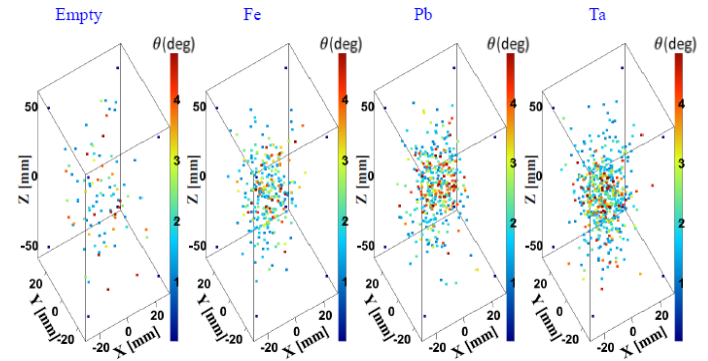

Figure 5: 3D reconstruction of the experimental data for the four scenarios. The color for each point represents the scattering angle in degree.

The plots in Fig. 6 show the 2D projections of thick slices in the x-y plane (top view) and y-z plane (side view) of the MTS volume for the four scenarios. For the top view, the MTS volume is divided into $2 \times 2 \times 20$ mm$^3$ voxels with the slice at z=0 shown. For the side view, the MTS volume is divided into $2 \times 2 \times 10$ mm$^3$ voxels with the slice at x=0 shown. For each voxel, we plot the mean scattering angle $<\theta>$ in degree for all those POCA "points" reconstructed inside that voxel. The white open squares, rectangles, and circles represent the actual dimensions and nominal locations of the targets within the MTS volume. We can clearly reconstruct, i.e. detect, and image these rather small targets. The voxels with high angles are located inside the targets, especially for the Ta and Pb targets. The circular Ta shape and the rectangular Pb shape as well as their dimensions are reproduced.

We run Monte Carlo (MC) simulation using the GEANT4 toolkit [13] for the four MTS scenarios with statistics equivalent to the data and compare the reconstruction of the scenarios for the simulation with our experimental data.

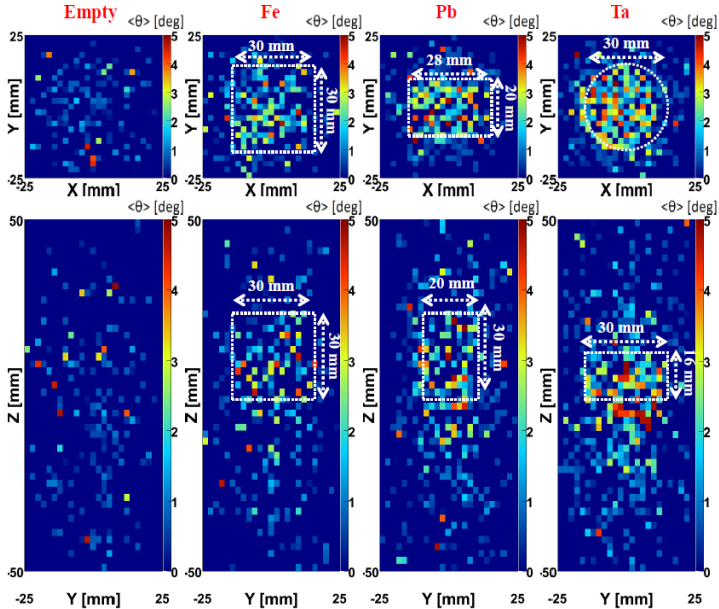

Figure 6: 2D reconstruction of experimental data for the four scenarios in X-Y slices (top) and Y-Z slices (bottom). The shading (color) of the voxels represents the mean value of the scattering angle in degree.

The plots in Fig. 7 show the top-view slices (x-y planes) at z=0 for data (top) and simulation (bottom) for both iron (Fe) and tantalum (Ta).

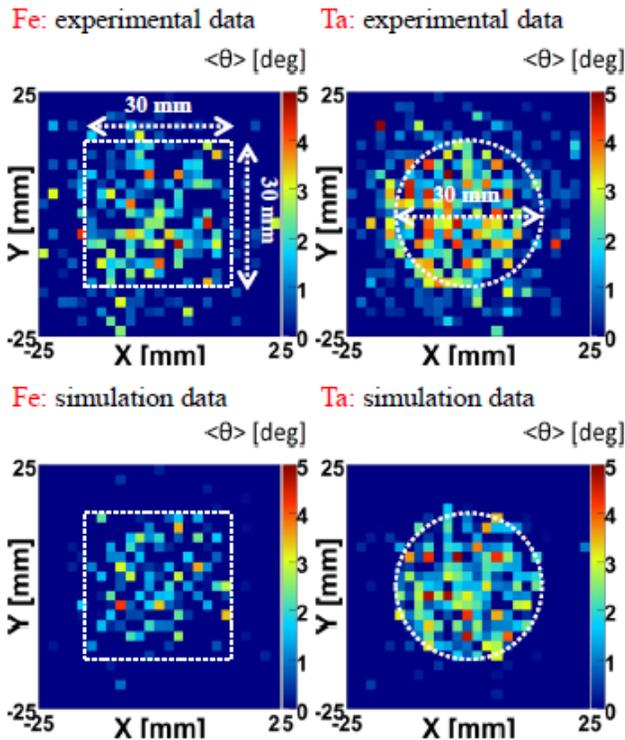

Figure 7: 2D reconstruction of the scenarios with iron (Fe) and tantalum (Ta): Experimental data (top) are compared with simulation (bottom). The shading (color) of the voxels represents the mean value of the scattering angle in degree.

There is a very good agreement between simulation and real experimental data. The background is more pronounced for the real data, which is explained by detector misalignment as well as by the fact that with the readout electronics used we were not operating at optimal spatial resolution (50-100 μm) that is expected for triple-GEM detectors. The measured mean angles are somewhat higher in data than in MC because the tracks from muons with small angles are more likely to reconstruct the "interaction" POCA point outside the voxel in which the actual interaction took place because of detector misalignment and poor spatial resolution as we previously demonstrated [5].

## 5. Conclusion and Future Plans

We have built and operated a first minimal MTS prototype using four GEM detectors and temporary electronics for reading out 768 channels (of ~15k total) as a first demonstration of using GEM detectors in a tracking station for muon tomography. Using several thousand cosmic ray muons recorded with the station, we are able to detect and image medium-Z and high-Z targets (Fe, Pb, Ta) with small volumes using our simple point-of-closest-approach reconstruction algorithm. This demonstrates that GEM-based muon tomography is in principle possible.

The next step is to fully instrument ten GEM detectors and mount them in a cubic-foot size MTS that also features side detectors. We are contributing to the effort made by the CERN RD51 collaboration [14] to develop a scalable readout system (SRS) [15] to read out all ~15k channels of this planned cubic-foot MTS prototype.

**Acknowledgment & Disclaimer**

We thank Leszek Ropelewski and the GDD group, Rui de Oliveira and the PCB production facility, and Miranda Van Stenis, all at CERN, for their help and technical support with the detector construction. We also thank Esther Ribas Ferrer, Fabien Jeanneau, Maxim Titov, from CEA Saclay (Paris, France) for lending us the Gassiplex FE cards and Theodoros Geralis from NCSR Demokritos (Athens, Greece) for getting us started on the LabView DAQ system. This material is based upon work supported in part by the U.S. Department of Homeland Security under Grant Award Number 2007-DN-077-ER0006-02. The views and conclusions contained in this document are those of the authors and should not be interpreted as necessarily representing the official policies, either expressed or implied, of the U.S. Department of Homeland Security.